\documentstyle [preprint, prd, aps]{revtex}

\def\gsim{\begin{array}{c} > \\ \sim \end{array}}
\def\lsim{\begin{array}{c} < \\ \sim \end{array}}

\begin{document}
\draft
\title{\large \bf From computation to black holes and space-time foam}
\author{\bf Y. Jack Ng\thanks{E-mail: yjng@physics.unc.edu}}
\address{Institute of Field Physics, Department of Physics and
Astronomy,\\
University of North Carolina, Chapel Hill, NC 27599-3255\\}
\maketitle

\begin{abstract}
We show that
quantum mechanics and general relativity limit the speed $\tilde{\nu}$ of
a \emph {simple} computer (such as a black hole) and its memory 
space $I$ 
to $\tilde{\nu}^2  I^{-1} \lsim t_P^{-2}$, where $t_P$ 
is the Planck time.  We also show that the life-time of 
a \emph {simple} clock and its
precision are similarly limited.  These bounds and
the holographic bound originate from the same
physics that governs the quantum fluctuations of space-time.
We further show that these physical
bounds are realized for black holes, yielding the correct Hawking black
hole lifetime, and that space-time
undergoes much larger quantum fluctuations than conventional
wisdom claims --- almost within range of detection with 
modern gravitational-wave interferometers.

\bigskip
PACS numbers: 04.70.Dy; 89.80.+h; 04.62.+v; 03.67.Lx

\end{abstract}

\newpage

The past few decades have witnessed amazing growth in the ability and speed 
with which computers can process information.  Quantum computation only adds 
to the prospect that this exponential
growth in information processing power will continue.  But it is natural
to ask whether this growth can go on indefinitely or whether there are
physical laws that impose  limitations to it. \cite{Ll,Pa} In this Letter
we show that indeed the laws of quantum mechanics and gravitation put
considerable bounds on computation.  In particular, the 
number $\tilde{\nu}$ of
operations per unit time, and the number $I$ of bits of information
in the memory space of a simple computer (``simple'' in the sense to be 
made precise below), are both limited by the input power
such that their product is bounded by a universal constant given 
by $\tilde{\nu}^2  I^{-1} \lsim
t_P^{-2}$, where $t_P = (\hbar G / c^5)^{1/2}$ is the Planck time formed
by the speed of light $c$, the quantum scale $\hbar$, and the
gravitational constant $G$.  Along the way, we also show that the
total running time $T$ over which a simple clock can remain accurate, and the 
smallest time interval $t$ that the clock is capable
of resolving, are bounded by $T \lsim t (t/t_P)^2$ .
Interestingly, these bounds are saturated for black holes.  So black holes,
in some sense,
may be regarded as the ultimate simple computers and 
ultimate simple clocks (though it may be
extremely difficult or even impossible to realize this technological
feat).  As a demonstration of the unity of physics, we show that the
physics that sets the limits to computation is precisely the physics that
governs the quantum fluctuations of space-time \cite{NvD1,Ka} which, as
pointed out recently,\cite{AC,Ah,NvD2} can plausibly be detected with
gravitational-wave interferometers such as LIGO/VIRGO and LISA through
future refinements.  Furthermore, the same physics underlies the holographic 
principle.  Thus the physics behind simple clocks, simple computers, black 
holes, space-time foam, and the holographic principle is inter-related.
It is this inter-relationship that we would like to emphasize in this Letter. 

The ingredients we use to derive the physical limits to computation are
the general principles of quantum mechanics and general relativity which 
should suffice for the physics in
the low-energy regime of quantum gravity in which we are interested.  (Thus, in what 
follows, it is understood that all the time intervals we are dealing 
with are much greater than the Planck time, and all the distances much 
larger than the Planck length ($ct_P$).)  Following
Wigner \cite{SW}, one can use quantum mechanics to set fundamental
limits on the mass $m$ of any system that serves as a time-registering
device.   Briefly, the argument goes as follows: If the clock has a linear
spread of $\delta R$, then its momentum uncertainty is $\hbar (\delta
R)^{-1}$.  After a time $\tau$, its position spread grows to $\delta
R(\tau) = \delta R + \hbar \tau m^{-1} (\delta R)^{-1}$ with the minimum
at $\delta R = (\hbar \tau / m)^{1/2}$.  
At the end of the total running time $T$, the linear spread can grow to
\begin{equation}
\delta R \gsim \left(\frac {\hbar T}{m}\right)^{1/2}.
\label{W1}
\end{equation}
But for the clock to give time to within accuracy $t$, it must have a small
enough spread in position, so small that the time at which a light quantum
strikes it (in order to read the time) can be determined within the
required accuracy $t$, thus $\delta R \lsim ct$.  In other words, we
require the wave packet of the center-of-mass of the clock be confined,
throughout the running time $T$, to a region of the size $ct$.  It follows
that, 
for a given $T$ and $t$, the lower bound on $m$ reads
\begin{equation}
m \gsim \frac {\hbar}{c^2 t} \left(\frac {T}{t}\right).
\label{W2}
\end{equation}
This limit is more restrictive than that given by Heisenberg's energy-time
uncertainty relation because it requires repeated measurement of time not
to introduce significant inaccuracies over the total running time $T$; this
requirement imposes a more severe limit on the clock mass than a simple
simultaneous measurement of both the time $t$ and the energy $m c^2$.

Next, as shown by the present author and van Dam\cite{NvD1}, one can 
supplement the
quantum mechanical relation Eq. (\ref{W2}) with a fundamental limit from
general relativity.  In essence one finds that the minimum time interval
that a clock can be used to measure is the light travel time across its
Schwarzschild radius.  The argument is quite simple.  Let the clock be a
simple light-clock consisting of two parallel mirrors (each of mass $m/2$)
between which bounces a beam of light.  On the one hand, for the clock to
be able to resolve time interval as small as $t$, the mirrors must be
separated by a distance $d$ with $d/c \lsim t$.  On the other hand,
$d$ is necessarily larger than the Schwarzschild radius $Gm/c^2$of the
mirrors so that the time registered by the clock can be read off 
at all.  From these two requirements, it follows that 
the upper bound on $m$ is given by\cite{NvD1}
\begin{equation}
t \gsim \frac {Gm}{c^3}.
\label{vD1}
\end{equation}
As clocks, black holes saturate this bound (more on this later).

One can now use Eq.(\ref{W2}) to obtain a bound on the speed of computation
of any information processor. \cite{Ba} The mean input power given by
$P = m c^2/T$ and the fastest possible processing frequency given by $\nu =
t^{-1}$ are bounded (via Eq.(\ref{W2})) as
\begin{equation}
\nu^2 = \frac {1}{t^2} \lsim \frac {mc^2}{\hbar T} = \frac {P}{\hbar}.
\label{B1}
\end{equation}
Thus power limits speed of computation.
 
Next, by substituting Eq. (\ref{W2}) into Eq. (\ref{vD1}) we can relate
T to t as
\begin{equation}
T \lsim t \left( \frac {t}{t_P} \right)^2.
\label{Tt}
\end{equation}
Thus the better precision a clock attains, i.e., the 
smaller t is, the shorter it
can keep accurate time, i.e., the smaller T is.  With the Planck time being
only about $10^{-43}$sec, this bound on $T$ is of no practical 
consequence (yet).  For example, a femtosecond ($10^{-15}$ sec)
precision yields the bound $T \lsim 10^{34}$ years.

Now it is time for us to make precise what we mean by the qualification 
``simple'' characterizing the simple clock and the simple computer.  
Alert readers may have already questioned the validity of Eq. (\ref{vD1}),
and accordingly also of the above T-t relation (Eq. (\ref{Tt})).  For
example, consider a large clock consisting of N identical small
clocks to keep time one after another. For large enough N, the T-t
relation and Eq. (\ref{vD1}) are violated for the large clock.  But 
note that this argument is not valid if
we consider only those clocks for which no
such separation of components is involved.  They are what we 
call \emph {simple} clocks.  The same qualification will be understood 
to apply to \emph {simple} computers.  The origin of this qualification can
be traced to the bound given by Eq. (\ref{vD1}).  Why should we be 
interested in simple clocks and simple computers?  For the simple reason 
that Nature makes use of them.  (This point will be made clear below 
when we derive 
the holographic principle and when we discuss the case of black holes.)

Let us use the T-t relation in Eq. (\ref {Tt}) to put a limit on the
memory space of a computer.  The point is that $T/t$, the maximum number
of steps of information processing, is, aside from factors like $ln 2$,
the amount of information $I$ that can be registered by the computer.
With the aid of Eq. (\ref{B1}), the T-t relation yields
\begin{equation}
I \sim \frac {T}{t} \lsim \frac {1}{(\nu t_P)^2} \sim \frac {\hbar}{P
t_P^2}.
\label{N1}
\end{equation}
While it is not too surprising that the input power $P$ limits the speed
of computation $\nu$ (as given by Eq. (\ref{B1})), it is less
expected that power also limits memory space of a computer in the way
given by Eq. (\ref{N1}).  We note that 
Eq. (\ref {W2}) and Eq. (\ref {vD1}) can also be 
used to give 
$I \nu \lsim m c^2 / \hbar$ and
$\nu \lsim \hbar / (t_P ^2 m c^2)$ respectively. 

More interestingly, Eq.(\ref{N1}) shows that the product of $I$ and $\nu
^2$ is bounded by a universal constant.  Now $\nu = t^{-1}$ 
should be identified as the clock rate of a computer, i.e., the
number of operations per bit per unit time.  It is related to the 
number $\tilde{\nu}$ of operations per unit time 
by $\tilde{\nu} = I \nu$.  Then it follows from Eq. (\ref{N1}) that
\begin{equation}
\frac {\tilde{\nu}^2}{I} = I \nu ^2 \lsim \frac {1}{t_P^2} \equiv \frac {c^5}{\hbar G} \sim 10^{86} /sec^2,
\label{N2}
\end{equation}
independent of the mass, size, and details of the simple computer. For the numerical
value in Eq. (\ref{N2}), we have used the speed of light in vacuum for $c$ in 
$t_P$.  This expression (valid for simple computers) 
links together our concepts of information, gravity,
and quantum uncertainty.  We will see below that nature seems to respect this 
bound which, in particular, is realized for black holes.  The restriction to 
simple computers is the price we have to pay for the universality of this
bound.  For comparison, current laptops
perform about $10^{10}$ operations per sec on $10^{10}$ bits, yielding 
$\tilde{\nu}^2 I^{-1} \sim 10^{10} / sec^2$. 

As intriguing as the physical limits to computation are, it is perhaps even
more amazing that the physics behind them is also what governs the quantum
fluctuations of space-time.  To see this, let us consider measuring the
distance $R \gg ct_P$ between two points.  We can put a clock at one of 
the points
and a mirror at the other point.  By sending a light signal from the clock
to the mirror in a timing experiment we can determine the distance.  But
the quantum uncertainty in the positions of the clock and the mirror
introduces an inaccuracy $\delta R$ in the distance measurement.  The same
argument used above to derive the T-t relation now yields a similar bound
for $\delta R$: 
\begin{equation}
\delta R \left ( \frac {\delta R}{c t_P}\right)^2 \gsim R,
\label{vD2}
\end{equation}
in a distance measurement.\cite{NvD1,EP}  In a time measurement, an analogous
bound is
given by Eq. (\ref{Tt}) with $T$ playing the role of the measured time and
$t$ the uncertainty.\cite{NvD1}  This limitation to space-time measurements can be
interpreted as resulting from quantum fluctuations of space-time itself.
In other words, at short distance scales, space-time is foamy.  Thus the
same physics underlies both the foaminess of space-time and the limits to
computation and clock precision.  Not surprisingly, these bounds have
the same form. 
It is remarkable that modern gravitational-wave interferometers, through
future
refinements, may reach displacement noise levels low enough to test this
space-time foam model,\cite{NvD1,Ka}
because the intrinsic foaminess of space-time
provides another source of noise in the interferometers that can be highly
constrained experimentally.\cite{AC,NvD2}  According to one estimate
\cite{AC,NvD2}, if Eq. (\ref{vD2}) is correct, the ``advanced phase" of LIGO
is expected to achieve a noise level low enough to probe 
$t_P$ down to $10^{-41} sec$, only about two orders
of magnitude from what we expect it ($t_P \sim 10^{-43} sec$) to be!

Furthermore, the same physics is behind the holographic principle, which
states that the number of degrees of freedom of a region of space is
bounded (not by the volume but) by the area of the region in Planck units.
\cite {tH}  To see this, consider a region of space with linear 
dimension R.  Conventional wisdom 
claims that the region can be partitioned into cubes
as small as $(c t_P)^3$.  It follows that the number of degrees of freedom
of the region is bounded by $(R/ct_P)^3$, i.e., the volume of the region in
Planck units.  But according
to Eq. (\ref {vD2}), the smallest cubes into which we can partition the
region cannot have a linear dimension smaller than $(Rc^2 t_P^2)^{1/3}$.
Therefore, the number of degrees of freedom of the region is bounded by
$[R/(R c^2 t_P^2)^{1/3}]^3$, i.e., the area of the region in Planck units,
as stipulated by the holographic principle.\cite{NvD2}  (A judicious 
application of Eq. (\ref{N1}) can also yield this result.)
Thus the holographic principle has its origin in
the quantum fluctuations of space-time.  Turning the argument around,
we believe the holographic principle alone suggests that
the quantum fluctuations of space-time are as given by Eq. (\ref{vD2}) 
($\delta R \gsim (R c^2 t_P^2)^{1/3}$)  and hence are much larger than what
conventional wisdom\cite{MTW} leads us to 
believe ($\delta R \gsim ct_P$).\cite{stf3}

Let us now ask for what kind of physical systems are the physical
limits listed above saturated (order-of-magnitude-wise).  We will show
that there is at
least one (and very likely only one) such system: the system of black
holes.
Since black holes have an entropy given by the event horizon area in
Planck units,\cite {Be} the holographic bound is obviously realized.
(Alternatively, we can show this by using Eq. (9) and Eq. (10) below.)
Next, consider a black hole as a clock, then it is reasonable to expect
that the maximum running time of this gravitational clock is given by the
Hawking black hole life-time
\begin{equation}
T \sim T_H \equiv \frac {G^2 m^3}{\hbar c^4},
\label{B2}
\end{equation}
and that the minimum interval that the black hole can be used to measure
is given by the light travel time across the black hole's horizon
\begin{equation}
t \sim \frac {Gm}{c^3}.
\label{B3}
\end{equation}
It is interesting that both Eq. (\ref{B2}) and Eq. (\ref{B3}) can
actually be derived by appealing to Wigner's two inequalities Eq. (\ref
{W1}) and Eq. (\ref{W2}) and using the Schwarzschild radius of the black
hole as the minimum clock size. \cite{Ba}  Better yet, if we use
Eq. (\ref{W2}) and Eq. (\ref{vD1}), we can immediately 
find that the bound on $T$ is given by the Hawking black hole
life-time.  Thus, if we had not known of black hole evaporation, this 
remarkable result would have implied that there is a maximum lifetime for
a black hole!  Now note that according to Eq.
(\ref {B3}),  the limit on $t$ as shown in Eq. (\ref{vD1}) is saturated
for a black hole.  Furthermore, using Eq. (\ref{B2}) and Eq. (\ref{B3}) one 
can easily show that the bound given by Eq. (\ref{W2}) is saturated.  It then 
follows that all the subsequent bounds (from Eq. (\ref{B1}) to 
Eq. (\ref{N2})) are saturated for black holes.  As a check, 
we can combine Eq. (\ref{B2}) and Eq. (\ref{B3})
to yield $T /t^3 \sim t_P^{-2}$  which saturates the T-t bound given in
Eq. (\ref {Tt}).  On the other hand, when a black hole is considered
as an
information processor with power $P = mc^2/T_H \sim \hbar c^6
/ G^2 m^2$ , we can use Eq. (\ref {B3}) to obtain $\nu^2 \sim P/\hbar$
which realizes the bound given by Eq. (\ref {B1}).  Eq. (\ref{B2}) and
Eq. (\ref {B3}) can also be used to yield $I \sim T/t \sim \hbar/Pt_P^2$
which saturates the bound given by Eq. (\ref {N1}).   Finally, with both
$\nu -$ and $I-$bounds saturated, the universal bound on computation 
given by Eq.
(\ref {N2}) is also saturated for black holes.  All these results reinforce
the conceptual importance of black holes as  
the simplest and most fundamental \cite {smf}
constructs of space-time, which set the universal limits
to computation, clock precision, and numbers of degrees of freedom. 
These properties of black holes lead us to believe that
their very existence lends support to the physical
bounds presented in this paper and the relatively large quantum
fluctuations of space-time given by Eq. (\ref{vD2}).  By the same token,
detection of the space-time foam (Eq. (\ref{vD2})) will be an indirect
verification of Hawking black hole evaporation.  From our perspective, 
black hole physics is intimately related to space-time foam physics.

Finally we note that, for a 1-kg black hole 
computer, Eq. (\ref{B2}) and 
Eq. (\ref{B3}) yield both $I = T/t \sim 10^{16}$ bits and 
$\tilde{\nu} = I \nu = T t^{-2} \sim 10^{51}/sec$.  These results agree with 
those given by Lloyd on the physical limits to computation
in Ref.\cite {Ll}.  

To summarize, we have shown that the laws of quantum mechanics and 
gravitation, 
which govern the quantum fluctuations of space-time,
also set physical bounds on computation and on the precision of clocks.
Power 
limits a simple computer's speed of computation $\tilde{\nu}$ and its 
memory space $I$.  Their product obeys the universal bound given by 
$\tilde{\nu}^2 I^{-1} \lsim t_P^{-2} \sim 10^{86}/sec^2$.  This bound
is realized for black holes.  The same physics underlies the 
holographic principle.  We have also argued that quantum fluctuations of 
space-time 
are actually much larger than what the folkfore suggests.  We urge the 
experimentalists, especially those in the gravitational-wave interferometer 
field,\cite{AC,NvD2,EP} to strive to detect them.

\bigskip

I thank C. Evans, C. Fuchs, S. Lloyd, E. Merzbacher, L. Ng, K. Rajagopal, 
G. 't Hooft, H. Tye, H. van Dam, R. Weiss, and D. Wineland for useful
discussions.  I thank S. Lloyd especially for pointing out my 
misidentification of computer speeds in the earlier versions of this
manuscript.  This work was supported in part by DOE under
\#DE-FG05-85ER-40219 and by the Bahnson Fund of University of North
Carolina at Chapel Hill.


\begin{references}
\bibitem{stf3}
There is another space-time foam model for which 
$\delta R \gsim (Rct_P)^{1/2}$.  See Ref. 5 and Ref. 7.  For this 
model, a calculation similar to that used in the text yields 
$I \nu \lsim t_P^{-1}$.  Unfortunately, this space-time 
foam model appears to be ruled out by the existing noise-level data 
obtained at the Caltech 40-m gravitational-wave interferometer.
\bibitem{smf}
The no-hair theorem for black holes already supports such a 
viewpoint.  In some sense, therefore, it is not surprising that 
black holes belong to the category of simple clocks and simple 
computers.  Note again the qualification ``simple''.
\end{references}
\end{document}